\documentclass[%
 reprint,
 amsmath,amssymb,
 aps,
 superscriptaddress,
]{revtex4-2}

\usepackage{graphicx}
\usepackage{dcolumn}
\usepackage{bm}
\usepackage{float}
\usepackage{tabularx}

\usepackage{caption}
\usepackage{subcaption}

\begin{document}
\preprint{APS/123-QED}

\title{Statistical and machine learning approaches for prediction of long-time excitation energy transfer dynamics}

\author{Kimara Naicker}
 \email{kimaranaicker@gmail.com}
\affiliation{%
 Quantum Research Group, School of Chemistry and Physics, University of KwaZulu-Natal, Durban, KwaZulu-Natal, 4001, South Africa 
}%
\affiliation{%
 National Institute for Theoretical and Computational Sciences (NITheCS), South Africa
}%
\author{Ilya Sinayskiy}
\affiliation{%
 Quantum Research Group, School of Chemistry and Physics, University of KwaZulu-Natal, Durban, KwaZulu-Natal, 4001, South Africa 
}%
\affiliation{%
 National Institute for Theoretical and Computational Sciences (NITheCS), South Africa
}%
\author{Francesco Petruccione}
\affiliation{%
 Quantum Research Group, School of Chemistry and Physics, University of KwaZulu-Natal, Durban, KwaZulu-Natal, 4001, South Africa 
}%
\affiliation{%
 National Institute for Theoretical and Computational Sciences (NITheCS), South Africa
}%
\affiliation{School of Data Science and Computational Thinking and Department of Physics, Stellenbosch University, Stellenbosch, 7600, South Africa}

\date{\today}

\begin{abstract}

One of the approaches used to solve for the dynamics of open quantum systems is the hierarchical equations of motion (HEOM). Although it is numerically exact, this method requires immense computational resources to solve. The objective here is to demonstrate whether models such as SARIMA, CatBoost, Prophet, convolutional and recurrent neural networks are able to bypass this requirement. We are able to show this successfully by first solving the HEOM to generate a data set of time series that depict the dissipative dynamics of excitation energy transfer in photosynthetic systems then, we use this data to test the models ability to predict the long-time dynamics when only the initial short-time dynamics is given. Our results suggest that the SARIMA model can serve as a computationally inexpensive yet accurate way to predict long-time dynamics. 

\end{abstract}

\maketitle

\section{\label{sec:level1}Introduction}

Time series analysis involves methods of analysing a series of data points that are indexed in time order. The objective of the analysis is to collect and study the past observations of a time series to develop an appropriate model which describes the inherent structure of the series. This model is then used to generate future values for the series, i.e. to make forecasts \cite{Bishop06}. In this work, the data being analysed is relevant to the dissipative dynamics of excitation energy transfer (EET) in systems similar to the photosynthetic open quantum system regime.

In some cases, information about the underlying dynamical correlations in open quantum systems can be encoded at the initial stages of their evolution. Therefore, it may be possible to obtain long-time dynamics of open quantum systems from the knowledge of their short-time evolution. This conjecture allows the bypass of the need for direct long-time simulations. The simulation of numerically exact methods to describe the dynamics of open quantum systems often require immense computational resources that scale exponentially with the size of the system under study, hence, it is desirable to develop an approach that can accurately predict long-time dynamics of open quantum systems along with eliminating the need for direct calculations to some extent.

Various numerical solutions for the dynamics of open quantum systems have been developed considering the complexity of system-bath interactions. The dynamics of an open quantum system that are dependent on the Hamiltonian of the system can be described through density matrix-based approaches in the Liouville space of the system. The numerically exact formalism adopted in this study is the hierarchical equations of motion (HEOM) developed by Tanimura and Kubo, and later adapted to biological light harvesting complexes by Ishizaki and Fleming \cite{Tanimura89, Tanimura20, ishizaki1}. Machine learning (ML) has been applied to this focus area in many relevant cases \cite{Lin21, Nelson22, ullah22, Rodriguez20, Herrera22, Naicker22}. L. E. Herrera \textit{et al.} conducted a comparative study where they benchmarked ML models based on their efficiency in predicting long-time dynamics of a two-level quantum system linearly coupled to harmonic bath \cite{Herrera22}.  

Successful time series forecasting depends on an appropriate model fitting. The development of efficient models to improve forecasting accuracy has evolved in literature. A comparison of the predictive capabilities of a standard statistical, an additive regression and a tree-based model against more structurally complex neural network models to simulate the open quantum system dynamics is carried out using Python. The first stage of the dynamics is obtained by solving the HEOM for a sufficiently large theoretical system, thereafter, we train and test suitable models to determine the validity of our approach. That is to predict a time series from that series past values efficiently. 

This paper contains several sections which are organized as follows: in Sec~\ref{sec:data} we describe the formalism used, Sec~\ref{sec:preprocess} describes the data pre-processing procedure, Sec~\ref{sec:method} covers the various time series models used, Sec~\ref{sec:results} presents our experimental forecasting results in terms of MSE obtained on relevant datasets and a brief conclusion of our work as well as the prospective future aim in this field.

\section{\label{sec:data}The theory and the data}

A time series is a sequential set of data points measured over successive times. It is mathematically defined as a set of vectors $x(t), t = 0, 1, 2, \dots$ where t represents the time elapsed \cite{Cochrane97}. The variable $x(t)$ can be treated as a random variable. The measurements taken during an event in a time series are arranged in chronological order.

Quantum systems faced in the real world are rarely entirely isolated, hence, it is important to consider the influence of the surrounding environment (bath) when studying the dynamical behaviour of a system. In the process of an open quantum system, such as a photosynthetic pigment-protein complex, evolving over time we can generate a set of time dependent observables that depict the coherent movement of electronic excitations through the system by solving the HEOM. This section describes the theoretical background used to generate the data sets used in the study.

The total Hamiltonian is composed of the Hamiltonian of the system, bath and system-bath interaction,
\begin{equation}
\label{eqn:Htot}
    \hat{H}_{Tot} = \hat{H}_{S} +\hat{H}_{B}+\hat{H}_{SB} .
\end{equation}
We focus on the simplest electronic energy transfer system, a dimer, where Hamiltonian of the system refers to the electronic states of a complex containing 2 pigments,
\begin{equation}
\label{eqn:Hs}
    \hat{H}_{S} = \sum_{j=1}^{2} |j\rangle\epsilon_j\langle j| + 
    J_{12} ( |1\rangle \langle 2| + |2\rangle \langle 1| ) ,
\end{equation}
where $\epsilon_{j}$ is the excited state energy of the $j$th site and $J_{12}$ denotes the electronic coupling between both sites. Here we consider that each pigment is coupled to a separate bath. The bath Hamiltonian represents the environmental phonons, 
\begin{equation}
    \hat{H}_{B} = \sum_{j=1}^2 \hat{H}_{B_{j}}, \quad \hat{H}_{B_{j}} = \sum_{\alpha} \hbar \omega_{j,\alpha} \bigg(\frac{\hat{p}_{j,\alpha}^2 + \hat{q}_{j,\alpha}^2}{2}\bigg) ,
\end{equation}
where $p$ is the conjugate momentum, $q$ is the dimensionless coordinate and $\omega_{j,\alpha}$ is the frequency of the $j$th site and $\alpha$th phonon mode, respectively. The last term of Eq.~\eqref{eqn:Htot} represents the fluctuations in the site energies caused by the phonon dynamics,
\begin{equation}
    \hat{H}_{SB} = \sum_{j=1}^{2} \hat{u}_j |j\rangle\ \langle j|  , \quad \hat{u}_j = \sum_{\alpha} g_{j,\alpha}\hat{q}_{j,\alpha} ,
\end{equation}
where $g_{j,\alpha}$ is the coupling constant between the $j$th site and $\alpha$th phonon mode.

The spectral density $\mathcal{J}_{j}(\omega)$ specifies the coupling of an electronic transition of the $j$th pigment to the environmental phonons through the reorganization energy $\lambda_{j}$ and the timescale of the phonon relaxation $\gamma_{j}$. Here it is expressed as the Ohmic spectral density with Lorentz-Drude cut-off, $\mathcal{J}_{j}(\omega) = 2\lambda_{j} \gamma_{j} \omega / (\omega^2 + \gamma_{j}^2)$. 

We focus on the application of this theory to EET at physiological temperatures of around 300 K, hence, the high-temperature condition characterized by $\hbar \gamma_{j} / k_{B}T \ll {1}$ is imposed and the following hierarchically coupled equations of motion are given \cite{ishizaki1},
\begin{eqnarray*}
    \frac{\partial}{\partial t} \hat{\sigma}^{(n_1, n_2)}(t) = - \left( i \hat{\mathcal{L}_e} + n_1\gamma_1 + n_2\gamma_2 \right) \hat{\sigma}^{(n_1, n_2)}(t) \nonumber
\end{eqnarray*}
\begin{eqnarray*}
    + \hat{\Phi}_1\hat{\sigma}^{(n_1+1, n_2)}(t) + n_1\hat{\Theta}_1\hat{\sigma}^{(n_1-1, n_2)}(t)
\end{eqnarray*}
\begin{equation}
\label{eqn:HEOM}
    + \hat{\Phi}_2\hat{\sigma}^{(n_1, n_2+1)}(t) + n_2\hat{\Theta}_2\hat{\sigma}^{(n_1, n_2-1)}(t) .
\end{equation}
In Eq.~\eqref{eqn:HEOM}, the element $\hat{\sigma}(\textbf 0, t)$ is identical to the reduced density operator $\hat{\rho}(t)$, while the rest are auxiliary density operators. The Liouvillian corresponding to the Hamiltonian $\hat{H}_{S}$ is denoted by $\hat{\mathcal{L}_e}$ and the relaxation operators $\hat{\Phi}_{j}$ and and $\hat{\Theta}_{j}$ are given by Eqs.~\eqref{eqn:Le}, \eqref{eqn:phi} and \eqref{eqn:theta},
\begin{equation}
\label{eqn:Le}
    \hat{\mathcal{L}_e} = [\hat{H}_{S}, \hat{\rho}_{S}], 
\end{equation}
\begin{equation}
\label{eqn:phi}
    \hat{\Phi}_j = i V_j^\times , \quad V_j^\times y = [V_j, y], 
\end{equation}
\begin{equation}
\label{eqn:theta}
    \hat{\Theta}_j = i \bigg( \frac {2\lambda_j}{\beta \hbar^2} V_j^\times -i\frac {\lambda_j}{\hbar} \gamma_j V_j^\circ \bigg) , \quad V_j^\circ y = \{V_j, y\} .
\end{equation}
Formally the hierarchy in Eq.~\eqref{eqn:HEOM} is infinite and cannot be numerically integrated. In order to make this problem tractable, the hierarchy can be terminated at a certain depth. There are several methods of doing so and in this work we have chosen the following termination condition following Ishizaki and Fleming \cite{ishizaki2}. For the integers $\textbf n = (n_1, n_2)$ and for characteristic frequency $\omega_{e}$ of $\hat{\mathcal{L}_e}$ where
\begin{equation}
\label{eqn:terminator}
    \mathcal{N} \equiv \sum_{j=1}^{2} n_{j} \gg \frac{\omega_{e}}{\mathrm{min}(\gamma_1, \gamma_2)} ,
\end{equation} 
Eq.~\eqref{eqn:HEOM} is replaced by
\begin{equation}
\label{eqn:terminatorterm}
    \frac{\partial}{\partial t} \hat{\sigma}(\textbf n, t) = -i \hat{\mathcal{L}_e} \hat{\sigma}(\textbf n, t) .
\end{equation} 

\begin{eqnarray*}
    \frac{\partial}{\partial t} \hat{\sigma}(\textbf n, t) = - \left( i \hat{\mathcal{L}_e} + \sum_{j=1}^{N} n_j\gamma_j \right) \hat{\sigma}(\textbf n, t) \nonumber \\
\end{eqnarray*}
\begin{eqnarray}
\label{eqn:HEOM N level}
    + \sum_{j=1}^{N} \bigg[ \hat{\Phi}_j\hat{\sigma}(\textbf n_{j+}, t) + n_j\hat{\Theta}_j\hat{\sigma}(\textbf n_{j-}, t) \bigg].
\end{eqnarray}
Eq.~\eqref{eqn:HEOM N level} is the general form of the reduced hierarchy equation Eq.~\eqref{eqn:HEOM}. The general form is employed in the following section as it is solved to generate the time series used in training and testing the models in the subsequent sections for systems containing more than two sites. 
\section{\label{sec:preprocess}Data pre-processing}

\begin{figure}[ht]
    \centering
    \includegraphics[width=\linewidth]{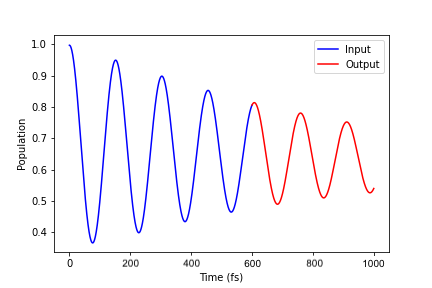}
    \caption{An example of a sequence generated by the HEOM for a dimer to be split to form the input and output data for the models.}
    \label{fig:fig1}
\end{figure}

Building a representative data set is an important first step in every machine learning project. Though the approach developed in this work can be generalized, we discuss the simplest electronic energy transfer system - a dimer (a spin-boson-type model \cite{Leggett87}) or two-level system as well as three- and four-level systems where linear chain configurations were imposed. The total system under study can be fully determined by the five independent energy scales: the site energy $\epsilon_{j}$, the coupling strength $J_{jk}$, the reorganization energy $\lambda$, the cut-off frequency $\omega_{c}$ and thermal energy $k_{B}T$ of the bath. In order to create a data set suitable for a framework aimed at predicting quantum dynamics in all physically realizable non-Markovian regimes, three parameters were extensively sampled while fixing two parameters: the cut-off frequency $\omega_{c} = 53 cm^{-1}$ and thermal energy $k_{B}T$ of the bath where $T = 300 K$. These are the typical parameters of photosynthetic EET \cite{Brixner05, Cho05} as seen in Table~\ref{tab:params}. 

\begin{table}[h]
    \begin{tabularx}{\columnwidth}{X|X|X}
        \hline
        Parameter       & Lower limit $cm^{-1}$ & Upper limit $cm^{-1}$ \\
        \hline
        $\epsilon_j$    & -100      & 100 \\
        $J_{jk}$        & -100      & 100 \\
        $\lambda$       & 1         & 100 \\
        \hline
    \end{tabularx}
    \caption{The data set containing time-evolved reduced density matrices is generated for all combinations of the following parameters: the site energy $\epsilon_{j}$, the coupling strength $J_{jk}$ and the reorganization energy $\lambda$.}
    \label{tab:params}
\end{table}

The HEOM method implemented in Python script is used to solve equation \eqref{eqn:HEOM}. The hierarchy truncation is set to 20 which is a sufficient depth based on the chosen cut-off frequency \cite{ishizaki2}. To make sense of time series data, it has to be collected over time in the same intervals. The total propagation time is set to 1.0 ps as is sufficient for observing coherent dynamics in photosynthetic EET \cite{ishizaki2}. For each of the 40000 samples, observables based on the diagonal elements of the time dependent density matrices i.e. the time evolution of the site populations are collected. Both of the generated observables are divided into shorter trajectories or sequences to test the capability of the models for varying output times to be predicted. 

One of our objectives is to determine the trade-off between how far ahead the model can forecast and the shortest input time required to maintain high accuracy in the forecast. As the total propagation time is fixed to 1.0 ps, the actual size of the dataset used varies depending on the lengths of the input and output times. We have performed a grid search in our model testing across varying lengths of input times up to 0.2 ps and output times between 0.01 ps and 0.6 ps. 

\begin{figure}[h]
\centering
  \includegraphics[width=\linewidth]{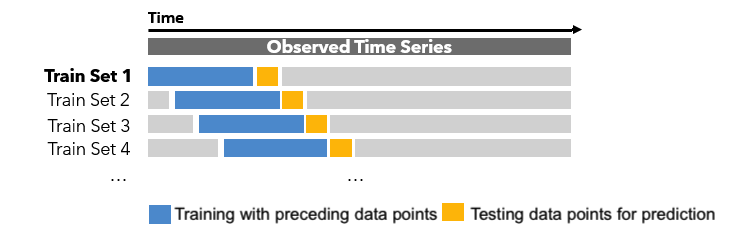}
\caption{The use of preceding time steps to predict the following time steps is referred to as the sliding window process. In the figure, the blue portion represents the points in the series that will be used for training and the yellow portion will be used in testing the models. Each training/ testing set differs as the "window" is moved forward over the entire time series by keeping the length of the portions fixed.}
\label{fig:sliding window}
\end{figure}

Before testing, each sample time series was split into multiple shorter slices. For example, a single series split to produce inputs that are 0.2 ps long and outputs 0.6 ps long would generate 1001 shorter sequences. This sliding window technique is pictured in Figure~\ref{fig:sliding window}

The data set is partitioned into a training set of 70\% of the data and a validation set of 10\% of the data. Additionally, 20\% of the data is held out during the training procedure and is used for testing.

\section{\label{sec:method}Methodology}

In practice, a suitable model is fitted to a given time series and the corresponding parameters of the underlying are estimated using the known data values. The procedure of fitting a time series to a proper model is known as Time Series Analysis. It comprises of methods that attempt to understand the nature of the series and is often useful for future forecasting and simulation. Past observations are collected and analyzed to build a suitable mathematical model which captures the underlying data generating process for the series. Then in time series forecasting, the future events are predicted using the model \cite{Zhang03, Zhang07}. 

Competent time series analysis remains dominated by traditional statistical methods as well as simpler machine learning techniques such as ensembles of trees and linear fits.

Machine learning is a sub-field of artificial intelligence. Deep learning is a sub-field of machine learning, and neural networks make up the backbone of deep learning algorithms \cite{LeCun15}. 
Machine learning can be broadly defined as computational methods using experience to improve performance or to make accurate predictions. Here, experience refers to the past information available to the learner. Machine learning algorithms can be broadly categorized as unsupervised or supervised by what kind of experience they are allowed to have during the learning process. Supervised learning algorithms experience a data set containing features but each example is also associated with a label or target. In supervised learning, the goal is to learn the relationship between a set of inputs and outputs.
Machine learning is intrinsically related to data analysis and statistics considering that the success of a learning algorithm largely depends on the data used. These techniques combine fundamental concepts in computer science with ideas from statistics, probability and optimization. Machine learning in time series analysis dates back many decades. A seminal paper from 1969, “The Combination of Forecasts,” analyzed the idea of combining forecasts rather than choosing a best one as a way to improve forecast performance. 

Inherently, ensemble methods have set the standard in many forecasting problems. Ensembling rejects the concept of a perfect or even significantly superior forecasting model relative to all possible models. 

We aim to estimate which of the models discussed in the following subsections can accurately predict the long-time dynamics of an open quantum system provided the preceding short-time dynamics of the system is known. 
We have employed the ETNA Time Series Library \cite{etna} for constructing and testing the classic SARIMA framework, the CatBoost model and Facebook's in-house model Prophet (specifically designed for learning from business time series) discussed in sections \ref{subsec:sarima}, \ref{subsec:catboost} and \ref{subsec:prophet}, respectively. ETNA is a user-friendly time series forecasting framework which includes built in toolkits for time series pre-processing, feature generation with a variety of predictive models. By using this package, the process of determining the parameters for the models is automated and optimized. In section \ref{subsec:neurals networks} we discuss neural network approaches for tackling sequential data problems.

\subsection{\label{subsec:sarima}SARIMA}

We consider conventional statistical methods developed specifically for time series data. The Auto-Regressive Integrated Moving Average (ARIMA) model and its variants account for the correlations that arise between data points in the same time series \cite{Box15}. This contrasts with the standard methods applied to cross-sectional data in which it is assumed that each data point is independent of the others in the sample. 
The Auto-Regressive (AR) model relies on the insight that the past predicts the future and so conjectures a time series process in which the value at a point in time t is a function of the series’ values at earlier points in time. 
A Moving Average (MA) model relies on a picture of a process in which the value at each point in time is a function of the past value error terms - each of which is independent from the others. 
Most time series models require the data to be stationary. This requirement is satisfied when the statistical properties such as mean, variance and covariance remain constant over time. Differencing is a means of removing trends and rendering a time series stationary. It is the process of converting a time series of values into a time series of changes in values over time. The model which combines these along with the added option of differencing is the ARIMA model which recognizes that a time series can have both underlying AR and MA model dynamics. 

The series obtained after differencing is the change between consecutive observations in the original series, it can be written as

\begin{equation}
\label{eqn:difference}
    y_{t}^\prime = y_{t} - y_{t-1},
\end{equation}

where $y_{t}$ is the $t^{th}$ term in the sequence of $T$ terms. Note that the series in Eq.~\eqref{eqn:difference} will have only $T-1$ values. 

The ARIMA model is specified in terms of the parameters $(p, d, q)$. The parameters refer to the auto-regressive, integrated and moving average parts of the data set, respectively. The p parameter is an integer that confirms how many lagged series are going to be used to forecast periods ahead. The d parameter indicates how many differencing orders are going to be used to make the series stationary, if necessary. The q parameter represents the number of lagged forecast error terms in the prediction equation. SARIMA is a seasonal ARIMA model and it is used for time series with seasonality. The uppercase notation $(P, D, Q)_m$ represent the seasonal components of the model where $m$ is the number of observations per year. 

We implemented the SARIMA framework from the publicly available ETNA package. It employs a grid search strategy that determines the optimal parameters for p, d, and q. The predictions on the test set are then obtained by using built-in functions and the results are presented in section \ref{sec:results}.

\subsection{\label{subsec:catboost}CatBoost}

CatBoost, short for category boosting, is an algorithm that is based on decision trees and gradient boosting. Decision trees are supervised learning method used for classification and regression \cite{Sammut10}. Gradient boosting is a process of constructing an ensemble predictor by performing gradient descent in a functional space \cite{Prokhorenkova19}. Boosting is a way of creating an ensemble of models. It begins by fitting an initial model to the data, thereafter another model is constructed that focuses on accurately predicting the cases where the initial model performed poorly. As this process of boosting is repeated, the ensemble is formed which is expected to be better than either model individually. Gradient boosting is taken a step further than this as the next model is combined with previous models to minimise the overall prediction error. The method involves setting the target outputs for the next model based on the gradient of the error with respect to the previous prediction, hence, each subsequent model takes a step in the direction that minimizes prediction error. 
Although the primary function of this model is for handling categorical data, we find it worth testing whether this model presents any advantages in forecasting. We tested the CatBoost model from the publicly available ETNA package. The results based on the predictions made are presented in section \ref{sec:results}.

\subsection{\label{subsec:prophet}Facebook Prophet}

Facebook's Prophet is an additive regression model designed for making forecasts for uni-variate time series datasets and to automatically find an optimal set of hyperparameters for the model with trends and seasonal structure by default. It belongs to the family of General Additive models (GAM) which fit a set of smooth functions that describe trend, seasonality and predictable special events or cycles to the data. The GAM is the sum of its smooth functions. A GAM treats a time series as a curve-fitting exercise.

Its core comprises of the sum of three functions of time plus an error term: growth or trend $g(t)$, seasonality $s(t)$, holidays $h(t)$ and error $e_t$. The growth function incorporates "changepoints", which are moments in the data where the data shifts direction, to model the overall trend of the data. The seasonality function is a Fourier Series as a function of time. Prophet automatically detects the Fourier order which is the optimal number of terms in the series. The holiday function allows Prophet to adjust accordingly to an anomaly that may change the forecast \cite{Taylor17}. The error term stands for random fluctuations that cannot be explained by the model.

We used the publicly available ETNA implementation of Prophet. The data had to be transformed for use of the model as the input data must contain two fields. Each data point in the time series had to be allocated a date (this must be a valid calendar date from which the holidays can be computed). We allocated a consecutive set of dates to each data point. The second field is the target variable which represents the value to be predicted. The results obtained by using built-in fitting and forecasting functions are presented in section \ref{sec:results}.

\subsection{\label{subsec:neurals networks}Neural networks}

Artificial neural network (ANN) is the broadest term used to classify a machine that mimics human intelligence. Similar to the human brain, ANNs attempt to recognize regularities and patterns in data, learn from experience and provide generalized results. They are data driven and self-adaptive in nature. It is not necessary to make any a priori assumption about the statistical distribution of the data; the desired model is adaptively formed based on the features presented from the data. This approach is useful for many practical situations, where no theoretical guidance may be available. Although the development of ANNs was mainly biologically motivated, they have been applied in various focus areas for forecasting and classification purposes.

Many of the steps of pre-processing data to fit a model’s assumptions are bypassed when neural networks are used - there is no requirement of stationarity, there is no need to develop the art and skill of picking parameters, such as assessing seasonality and order of a seasonal ARIMA model and there is no need to develop a hypothesis about the underlying dynamics of a system, as is helpful with state space modelling. However, this category of algorithms does impose its own pre-processing requirements which we will discuss later.

The first concepts in deep learning brought about three main choices of architecture depending on the data type being handled: Artificial Neural Networks (ANNs) mainly for classification and regression problems; Convolutional Neural Networks (CNN)-based for spatial data (such as images data) and Recurrent Neural Networks (RNN)-based for sequential data (such as time series data). CNNs and RNNs incorporate feature engineering into their framework and eliminate any need to do so manually. This is seen in their capacity to extract features and create informative representations of time series automatically. 

The neural network models discussed in sections \ref{subsubsec:lstm} and \ref{subsubsec:cnn} have been built, trained and tested using the Scikit-learn \cite{scikit-learn} and TensorFlow \cite{tensorflow2015-whitepaper} Python packages. 

\subsubsection{\label{subsubsec:lstm}LSTM}

\begin{figure}[t]
\centering
  \includegraphics[width=\linewidth]{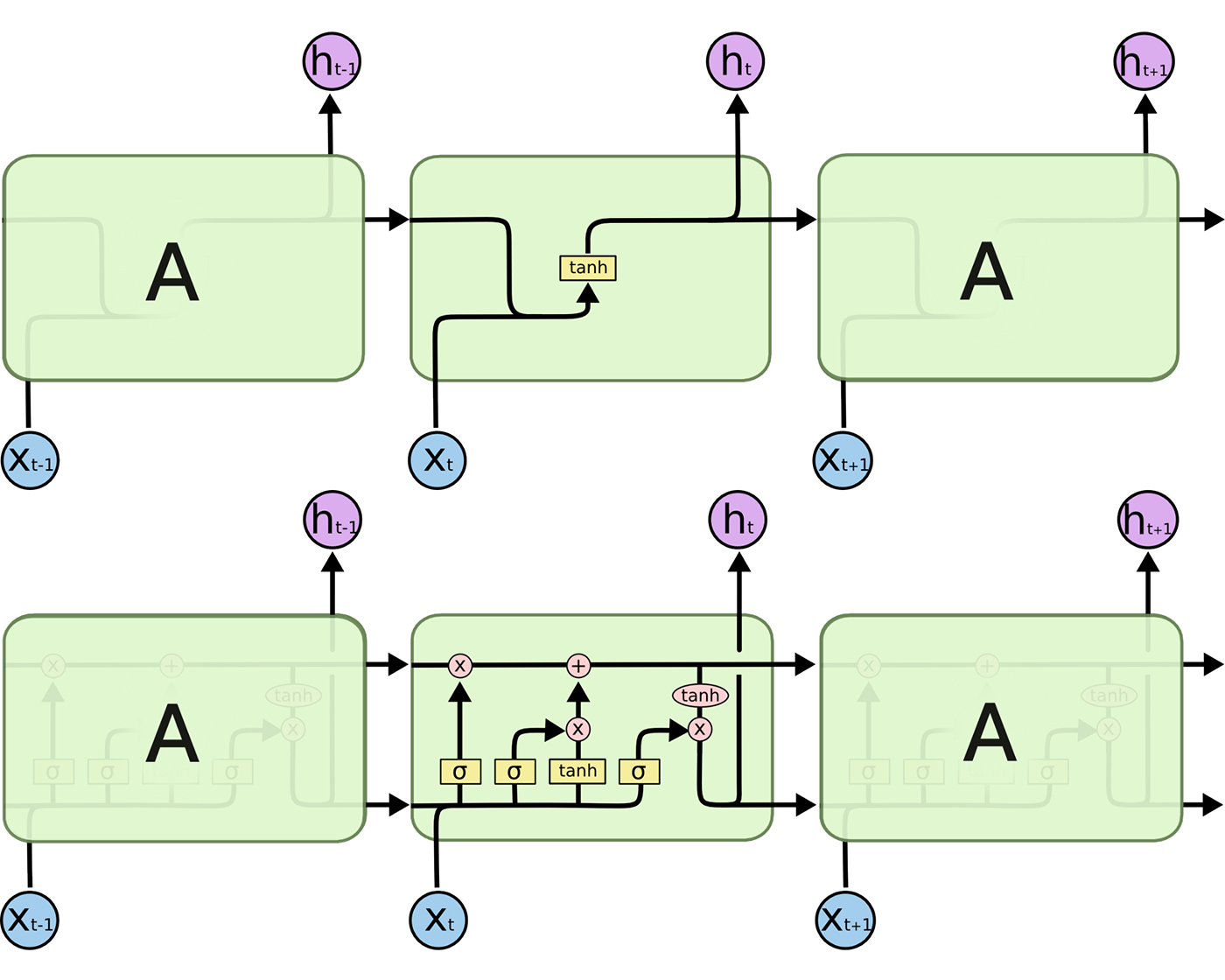}
\caption{RNN cell structure (top) vs. LSTM cell structure (bottom). Each line carries an entire vector, from the output of one node to the inputs of others. The yellow circles represent pointwise operations, like vector addition, while the orange boxes are learned neural network layers. Lines merging denote concatenation, while a line forking denote its content being copied and the copies going to different locations.}
\label{fig:rnnVsLstm}
\end{figure}

At a basic level, a neural network — of which recurrent neural networks (RNNs) are one type, among other types such as convolutional networks (discussed in section \ref{subsubsec:cnn}) — is composed of three primary components: the input layer, the hidden layers and the output layer. Each layer consists of nodes or neurons. Feed-forward neural networks (FFNNs), the original single-layer perceptron, developed in 1958 is the precursor to recurrent neural networks. In FFNNs, the information flows in only one direction: from the input layer, through the hidden layers, to the output layer but never backwards in feedback loops.

A recurrent neural network, by contrast, retains a memory of what it has processed. Recurrent neural networks are a broad class of networks specifically designed for processing sequential data. They are attractive to us in this study because of their characteristic capabilities of analysing temporal data as they can retain their state from one iteration to the next by using their own output as input for the next step. In other words, RNNs are a type of neural network that makes recurrent connections by going through temporal feedback loops, thus they can learn from previous iterations during its training. The loops make it a recurrent network.

The hidden layers are placed between the input and the output layer. In an RNN, an output is produced but is also fed back through \textit{backpropagation} as an input for training the hidden layer on the next observation. RNNs carry out the training process by adjusting the weights throughout the neural network. The network recalibrates the weights for both the current and the previous inputs, multiplies the vector of input values with the vector of new weights - this step either enhances or diminishes the importance of each input with respect to the training goal of lowering the prediction error - and lastly, passes the vector of results on as an input to the next layer. By adapting the weights, the hidden layer incrementally derives a function which transforms the input values to output values that approximate the actual observations in the training dataset. However, the function that maps the inputs to the outputs is not expressed as a closed-form equation i.e. it remains hidden.

We will discuss the main building blocks of RNNs. All RNNs have the form of a chain of repeating modules of neural network layers. In standard RNNs, this repeating module will have a simple structure. Within hidden layers, the receiving node calculates a weighted sum of the inputs it receives (which are the output of the preceding layer and the new input) to calculate the total input for the activation (transfer) function. The activation function determines how much the node will contribute to the next layer which is what the output represents. Among the types of frequently used activation functions are the sigmoid function, hyperbolic tangent function (tanh), the step function and the ReLU (rectified linear unit) function.

When the network generates prediction values after a forward pass, the prediction error is also computed which quantifies the deviation from the training dataset through a cost function (also referred to as a loss, error or objective function). The network aims to minimize the error by adjusting its internal weights during training. Backpropagation calculates the partial derivatives of the error with respect to the weights. In each iteration, the RNN recalibrates the weights, up or down, based on the partial derivatives. 

The concept of gradient descent refers to the search for a global minimum by evaluating the partial derivatives. The partial derivative with respect to a certain weight reveals how that particular weight contributes towards the total error. The network varies a single weight and records its effect on the total error to obtain its gradient. The repeated adjustment of the weights, down the descent towards a minimal error, will move the model towards an incrementally reduced prediction error. This is computationally intensive hence, the long training phases a RNN typically requires. Gradient descent denotes the search for the global minimum, the set of weights that will minimize the total error. 

The RNN updates the old weights by subtracting from them a fraction of their respective gradients. The fraction represents the learning rate, a value above 0 and up to 1 i.e. new weight = old weight — gradient * learning rate. A higher learning rate can speed up the training process of the RNN although, it can also cause overshooting which may prevent the network from settling on a minimal total error. The goal of the RNN is to minimize the cost function. Prediction accuracy metrics such as the mean squared error MSE or root mean squared error RMSE can serve as cost functions.

An epoch represents the feeding of the entire training dataset through the network, consisting of one forward and one backward pass. The number of epochs will determine the trade-off between the time required to train the RNN and its accuracy. 

A simple recurrent neural network works well only for a short-term memory. In practice, RNNs suffer from a fundamental problem where they are sometimes unable to capture longer time dependencies in the data. The problem was explored in depth by Hochreiter \cite{Hochreiter91} and Bengio \textit{et al.} \cite{Bengio94}, who found fundamental justifications for why this caveat exists.

Long Short-Term Memory networks (LSTMs) are a variation of RNNs explicitly designed to avoid the long-term dependency problem. They were introduced by Hochreiter and Schmidhuber \cite{Hochreiter97}. LSTMs process elements one at a time with memory thus, they are perceived to be more suitable for extracting long range temporal dependencies. Prior to LSTMs, RNNs were "forgetful". Meaning that they could retain a memory although, only pertaining to the process steps in their immediate past. The LSTM, by contrast, introduces loops that can generate long-duration gradients. It can hold on to long-term patterns it discovers while going through its loops. At each time step, it can collate three pieces of information: the current input data, the short-term memory it receives from the preceding cell i.e. the hidden state and the long-term memory from more remote cells (the so-called cell state), from which the RNN cell produces a new hidden state. LSTMs also have a chain like structure, however, the repeating module has a different and more complex structure as seen in Figure \ref{fig:rnnVsLstm}. 

The long-duration gradients resolved a problem termed vanishing gradient descent which occurs when the model stops learning because the gradient’s slope becomes too shallow for the search to further improve the weights. This can occur when many of the values involved in repeated gradient calculations are smaller than 1. On the contrary, exploding gradients arises when many values exceed 1 in the repeated matrix multiplications the RNN carries out. The vanishing gradient problem limits an RNN’s memory to short-term dependencies whereas the LSTM’s architecture keeps the gradients steep enough so that the search does not get "stuck". 

A cell in the LSTM is said to be “gated”. The cell selectively adds or removes information through the gates as it determines how much incoming information is captured and how much of it is retained. The model can decide whether it opens an input gate to store information, reject and delete it from long-term memory (forget gate) or passes the information on to the next layer (output gate). The RNN is able to make these decisions based on the weights it learns to assign in the process of minimising the error. The gates carry out matrix multiplications between the information values they receive as their current inputs, from short-term or long-term memory. Over time, the LSTM learns which information pieces are effective in reducing the prediction error. It will open and close gates accordingly by assigning higher or lower weights between 0 and 1 to the information values. Through its loops, it will let the useful values, with higher weights, pass through the output gate to form a new short-term memory and it will discard the low-weighted values.

The number of neurons and layers are user definable. Although, decreasing the number of neurons and hidden layers can improve the computation time per iteration, this comes at the cost of accuracy. The model also has an added caveat of either underfitting or overfitting. In the case of overfittng, the model will perform well on training data where as this performance will deteriorate on unseen test data. In order to develop a reliable mode, the model architecture needs to be determined by optimization possibly by a trial-and-error procedure. Many recent works have exhibited the capability of neural networks, specifically recurrent neural networks (RNNs), for sequential data analysis \cite{Bandyopadhyay18, Lara21, Lim21, Sezer20, Fawaz19}. 

Other RNN variants — and even other types of LSTMs — exist such as gated recurrent unit networks, bidirectional RNNs and convolutional RNNs which may be used in further testing and implementation. It is also important to note that studies find that the RNN variants do not consistently outperform one another, rather performance is reliant on the data to some extent \cite{Greff15}. There does not seem to be the best RNN variant.

The training and testing procedure for LSTMs is straightforward through the Scikit-learn package. Once the data is loaded in, pre-processed and split into training and validation datasets, it can be fit with the model. The main set of hyperparameters that are needed to set up the model before fitting are the number of input nodes, hidden layers and their nodes, output nodes, number of epochs and the learning rate. Their optimal settings are not known a priori, rather they are problem specific. The fitting or training process is time-consuming and dependent on the computer’s processor performance. Once the model is fit after training and validation, it can be tested on the test dataset. The results of this stage of testing of our model is presented in the results of section \ref{sec:results}.

\subsubsection{\label{subsubsec:cnn}CNN}

It has been shown that using CNNs for time series classification has several important advantages over other methods. They are highly noise-resistant models and they can extract informative features which are independent of time \cite{LeCun95}.

\begin{figure}[t]
    \centering
    \includegraphics[scale=0.3]{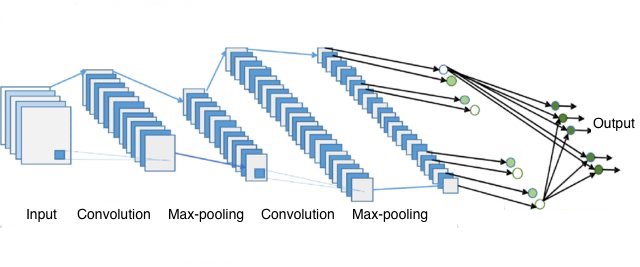}
    \caption{Vanilla convolutional neural network consisting of convolutional and max-pooling layers fed into a fully-connected network.}
    \label{fig: cnn}
\end{figure}

Consider a time series of length n and width k. The length is the number of time steps and the width is the number of variables in a multi-variate time series. Generally, the first layer in a CNN network is the convolutional layer. This is the core building block and does most of the computational heavy lifting. The convolution kernels have the same width as the time series and their length can vary. The kernel moves in one direction from the beginning of a time series towards its end, performing the convolution operation. The elements of the kernel get multiplied by the corresponding elements of the time series that they cover at a given point (referred to as the receptive field). The results of the multiplication are added together and a non-linear activation function is applied to the value in the activation layer. The most frequently used non-linearity for CNNs is the rectified non-linear unit (ReLU) function which combats the vanishing gradient problem occurring in the sigmoid function. The resulting value becomes an element of a new “filtered” uni-variate time series. The kernel then moves forward along the time series to compute the next value. The number of new “filtered” time series is equivalent to the number of convolution kernels. Depending on the length of the kernel, different characteristics of the initial time series get captured in each of the new filtered series. A representation of this layer can be seen in Figure~\ref{fig: cnn}. 

Next is the pooling layer. This involves the downsampling of features by applying the max-pooling to each of the filtered time series vectors, the largest value is taken from each vector based on the specified stride and size of the window. A new vector is formed from these values and this vector of maximums is the final feature vector that can be used as an input to a regular fully connected layer. Lastly is the fully connected layer, this involves flattening. In this layer, the entire pooled matrix is transformed into a single column which is then fed to the artificial neural network layers for processing.

There are several significant differences in the working nature of CNNs versus RNNs that make each or the other better suited for a task. CNNs use convolution operations that can handle spatial information available in images. CNNs are computationally cheaper than RNNs as a CNN learns in batches while RNNs train sequentially, as a consequence, a caveat of using an RNN is that parallelization cannot be done. Unlike RNNs, CNNs learn patterns within a specified range of steps or time window without the assumption that history is complete. This property can make CNNs attractive for working with missing data. Data pre-processed for CNNs can be shuffled which allows the CNN to interpret the data from a broader perspective to a certain extent. RNN models are limited to learning from data in preceding time steps only. For the case of training a model such that it is dependent on the history of the data and/ or such that it can handle varying sizes of input and outputs, RNNs are more suitable. 

Similarly to RNNs, the training and testing procedure for CNNs is straightforward through the Scikit-learn package. Once the data is loaded in, pre-processed and split into training and validation datasets, it can be fit with the model. The main set of hyperparameters that are needed to set up the model before fitting are different for each of the three types of layers. For the convolutional layer, the number and size of kernels and the activation function. The pooling layer requires the stride and size of the window and the fully connected layers depend on the number of nodes and activation function. Once the model is fit after training and validation, it can be tested on the test dataset. In terms of CNNs, it is worth mentioning that the proposed method is not the only one that exists. There are ways of presenting time series in the form of images to which a regular 2D convolution can be applied. The results of this stage of testing of our model is presented in the results of section \ref{sec:results}.

\section{\label{sec:results}Results \& Conclusion}

\begin{table*}[ht]
    \begin{tabularx}{\textwidth}{X|X|X|X|X|X|X}
        \hline
        Model       & \multicolumn{2}{c}{2 level} & \multicolumn{2}{c}{3 level} & \multicolumn2{c}{4 level}\\
        \hline
        & MSE $(10^{-3})$ & Time to solve ($s$) & MSE $(10^{-3})$ & Time to solve ($s$)
         & MSE $(10^{-3})$ & Time to solve ($s$) \\
        \hline 
        SARIMA      & 0.3116 $\pm$ 0.0011     & 0.10  
                    & 0.3193 $\pm$ 0.0056     & 0.12     
                    & 0.3491 $\pm$ 0.0002     & 0.12 \\
        CatBoost    & 3.5634 $\pm$ 0.0155     & 4         
                    & 2.1060 $\pm$ 0.0143     & 5        
                    & 2.7800 $\pm$ 0.0244     & 7       \\
        Prophet     & 0.1281 $\pm$ 0.0099      & 23      
                    & 0.1906 $\pm$ 0.0100      & 26      
                    & 0.1821 $\pm$ 0.0140      & 27 \\
        CNN         & 1.3012 $\pm$ 0.0343      & 835  
                    & 12.616 $\pm$ 0.0254     & 900 
                    & 16.010 $\pm$ 0.0233     & 890 \\
        LSTM        & 0.0461 $\pm$ 0.0090      & 7200
                    & 0.0923 $\pm$ 0.0109      & 7400
                    & 0.1204 $\pm$ 0.0100      & 7500\\
        \hline
    \end{tabularx}
    \caption{The MSE values and time to train/ test for each model to predict 100 time steps ahead i.e. 0.02 fs.}
    \label{tab:results}
\end{table*}

\begin{table*}[ht]
    \begin{tabularx}{\textwidth}{X|X|X|X|X|X|X}
        \hline
        Model       & \multicolumn{2}{c}{2 level} & \multicolumn{2}{c}{3 level} & \multicolumn{2}{c}{4 level}\\
        \hline
        & MSE $(10^{-3})$ & Time to solve ($s$) & MSE $(10^{-3})$ & Time to solve ($s$)
         & MSE $(10^{-3})$ & Time to solve ($s$)\\
        \hline
        SARIMA      & 9.7242 $\pm$ 0.0121 & 0.27   
                    & 12.448 $\pm$ 0.0021 & 0.31 
                    & 11.306 $\pm$ 0.0123 & 0.37 \\
        CatBoost    & 13.596 $\pm$ 0.1098  & 7  
                    & 19.558 $\pm$ 0.1138 & 9 
                    & 14.855 $\pm$ 0.0198 & 8 \\
        Prophet     & 236.66 $\pm$ 0.3345 & 24
                    & 307.47 $\pm$ 0.4341 & 22
                    & 238.76 $\pm$ 0.3971 & 30\\
        CNN         & 22.681 $\pm$ 0.1143 & 4200 
                    & 65.123 $\pm$ 0.1909  & 3990
                    & 66.321 $\pm$ 0.0967  & 4300\\
        \hline
    \end{tabularx}
    \caption{The MSE values and time to train/ test for each model to predict 3000 time steps ahead i.e. 0.6 fs.}
    \label{tab:results2}
\end{table*}

\begin{figure}[t]
    \centering
    \includegraphics[width=\linewidth]{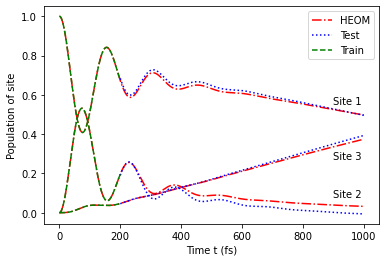}
    \caption{Comparison of HEOM results (red) with predictions made by SARIMA (blue) based on short-time training data (green) for sites 1-3 of the well-known photosynthetic seven-site pigment protein Fenna–Matthews–Olson complex.}
    \label{fig: FMOtest}
\end{figure}

In this section we report results obtained with the five different approaches to forecasting – namely SARIMA, CatBoost, Prophet, CNN and LSTM. All training and testing of models was carried out using a workstation with the following specifications: 2x Intel Xeon E5-2640 2.0GHz, 128 GB RAM. Maintaining consistency in the machine used is crucial as we are comparing the time required by each model as well. It is worth noting the difference in meaning of training for the neural networks versus the rest of the models. In the case of neural networks, in the training stage the model is trained on \textit{all the samples} in the training dataset. The weights and biases of the models are adjusted in each epoch until the minimum MSE is achieved. Only after this is the model used for predictions on the test dataset - it is the time taken to train and test the CNN and LSTM models and the MSE achieved on the test dataset that reported here. On the contrary, SARIMA, CatBoost and Prophet do not perform batch training and testing. The models train on a single sample time series and make predictions based on this single sample training only. Hence, it is the average MSE and time taken \textit{per sample} that is reported in this case. 

The accuracy of a models predictions depend on the length of the output time to be predicted as can be expected. Tables~\ref{tab:results} and \ref{tab:results2} capture the mean squared error results of each model for varying forecasting lengths. We choose an input time of 0.2 ps in all calculations reported in this study as it provides the best compromise between accuracy and computational cost. 

The mean-squared error (MSE) provides a direct quantitative check of the extent to which the predicted response value for a given observation is close to the true response value for that observation. This measure works well in ensuring that our trained model has no outlier predictions with large errors since it allocates larger weight to these errors due to the squaring part of the function. To calculate the MSE as given in Eq.~\eqref{eqn:MSE}, the difference between the models predictions $y_{i}$ and the ground truth $f_{i}$ are squared then, the average across the whole dataset is taken. An ideal MSE value is 0.0, which means that all predicted values matched the expected values exactly.

\begin{equation}
\label{eqn:MSE}    
    MSE = \sum_{i}^{N} \frac{(y_i - f_i)^2}{N}
\end{equation}

MSE, which squares the prediction errors, penalizes larger errors more than mean absolute percentage error (MAPE) does. Bias arises when the distribution of residuals is left-skewed or right-skewed. The mean will lie above or below the median. A forecast that minimizes the MSE will exhibit less bias.

We emphasize that only the short-time initial dynamical information is required - 1000 time steps or 0.2 ps. The rest of the time evolution is simulated by reconstructing the observable based on predictions for each time step beyond the initial input time. Thus, if a single-step prediction error is not sufficiently small then the error will rapidly accumulate resulting in a deterioration of the accuracy. As seen in the reported results, all models are able to reproduce long-time dynamics nearly exactly and irrespective of the dynamical regime. The LSTM model was excluded from the training and testing at the longer time due to its exhaustive computational requirements i.e. the trade-off between accuracy and time did not seem valid here.

With regards to hyperparameter settings, we have limited the learning rate, 0.001 and the epochs, 300, in our setup of the LSTM and CNN models. A tuning algorithm could tweak them while running the fitting process to try to achieve an even lower MSE. On the other hand, this particular dataset of time series is not greatly complex as it is purely dissipative. It does not seem likely that time-intensive tuning efforts would reduce the MSE much further.

The ability of each model to learn the fundamental properties of the density matrix is tested on the unit trace and positive semi-definite properties. We emphasize that these properties are not enforced during the training. We found that all models were able to learn both properties, however, the SARIMA model proved to be the fastest, most efficient and most accurate.
An observation to note is that each model produced roughly the same results for each level of systems. It is also evident how the complexity of the HEOM increases with increasing system size i.e. more computational power and time required to explicitly solve the set of equations. The importance of our results is that these machine learning models provide a cheaper yet still greatly efficient alternative to determining the long-time dynamics of these systems. This could be widely used depending on the level of accuracy required for a particular use case.
In Figure~\ref{fig: FMOtest}, we present the results from testing the predictive capability of the SARIMA model on the Fenna–Matthews–Olson (FMO) protein of green sulfur bacteria \cite{Fenna75}. The time evolution of the population of the first three sites of the complex are used for training and testing where site one is the initial excited state for numerical calculations. The FMO complex was one of the first systems of its kind in which long-lived quantum coherence was observed which motivated the development of new a quantum dynamic equation for excitation energy transfer i.e. HEOM, hence, the relevance and importance of testing our model on this data set.
Furthermore and as anticipated during testing, the models performed slightly worse in the region of small reorganization energy and low temperature which is the regime of weakly damped coherent oscillatory dynamics. This is largely due to the input data being almost indistinguishable in this regime. 
In summary, we demonstrate that SARIMA, CatBoost, Prophet, convolutional and recurrent neural networks models can predict the long-time dynamics of an open quantum system provided the preceding short-time dynamics of the system is known. However, owing to its efficiency with regards to computational power required and time required to conduct simulations plus, low MSEs achieved, the SARIMA appears to be the model that satisfies our criteria. The model has been trained on a data sets relevant to photosynthetic excitation energy transfer and we aim to scale and deploy it to investigate long-lasting quantum coherent phenomena observed in larger light-harvesting complexes. The approach is a practical tool in terms of reducing the required computational resources for long-time simulations whilst maintaining high accuracy. A photosynthetic EET was chosen as an example, however, the same approach can be used to study other phenomena provided the relevant data sets are available.

\section*{\label{sec:acknowledgement}Acknowledgements}

This work is based upon research supported by the South African Research Chair Initiative of the Department of Science and Innovation and the National Research Foundation of the Republic of South Africa. Support from the NICIS (National Integrated Cyber Infrastructure System) e-research grant QICSA is kindly acknowledged.

\nocite{*}
\bibliography{manuscript}

\pagebreak

\clearpage

\end{document}